\documentclass[aps,twocolumn,prb,superscriptaddress,showpacs]{revtex4}

\usepackage{graphicx}

\newcommand {\qEA}	{q_{\mathrm{EA}}}
\newcommand {\hp}	{\Delta h}
\newcommand {\tw}	{t_\mathrm{w}}
\newcommand {\Tc}	{T_\mathrm{c}}
\newcommand {\hc}	{h_\mathrm{c}}

\begin{document}

\title{Off-equilibrium fluctuation-dissipation relations\\ in the
3\boldmath{$d$} Ising Spin Glass in a magnetic field.}

\author{A. Cruz} 
\email{cruz@sol.unizar.es}
\affiliation{Depto. de F{\'\i}sica Te\'orica,
Facultad de Ciencias, Universidad de Zaragoza, 50009 Zaragoza, Spain.}
\affiliation{Instituto de Biocomputaci\'on (BIFI), Universidad de Zaragoza.}
\author{L. A. Fern\'andez} 
\email{laf@lattice.fis.ucm.es}
\affiliation{Depto. de F{\'\i}sica
Te\'orica, Facultad de CC. F{\'\i}sicas, Universidad Complutense de
Madrid, 28040 Madrid, Spain.}  
\affiliation{Instituto de Biocomputaci\'on (BIFI), Universidad de Zaragoza.}
\author{S. Jim\'enez} 
\email{sergio@rtn.unizar.es}
\affiliation{Depto. de F{\'\i}sica Te\'orica,
Facultad de Ciencias, Universidad de Zaragoza, 50009 Zaragoza, Spain.}
\affiliation{Instituto de Biocomputaci\'on (BIFI), Universidad de Zaragoza.}
\author{J. J. Ruiz-Lorenzo} 
\email{ruiz@unex.es}
\affiliation{Depto. de F\'{\i}sica,
Facultad de Ciencias, Universidad de Extremadura, 06071 Badajoz,
Spain.}  
\affiliation{Instituto de Biocomputaci\'on (BIFI), Universidad de Zaragoza.}
\author{A. Taranc\'on} 
\email{tarancon@sol.unizar.es}
\affiliation{Depto. de F{\'\i}sica Te\'orica, 
Facultad de Ciencias, Universidad de Zaragoza, 50009 Zaragoza, Spain.}
\affiliation{Instituto de Biocomputaci\'on (BIFI), Universidad de Zaragoza.}
\date{March 5, 2003}

\begin{abstract}

We study the fluctuation-dissipation relations for a three dimensional
Ising spin glass in a magnetic field both in the high temperature
phase as well as in the low temperature one. In the region of times
simulated we have found that our results support a picture of the low
temperature phase with broken replica symmetry, but a droplet behavior
can not be completely excluded.
 
\end{abstract}


\maketitle

\section{Introduction}

The understanding of the behavior of a spin glass in a magnetic field
is a challenging issue from both  experimental and theoretical sides.

In the theoretical side, there are two competing theories. In the
droplet model the spin glass phase is unstable (for any amount of
magnetic field) and so there is no phase transition: there is only a
pure state describing all the Ising spin glass in a magnetic field
.\cite{DROPLET} On the other side, Mean Field (MF) predicts a phase
transition between two phases. The first one is characterized by one
pure state, and the low temperature phase is described by a countable
number of pure states. In the Mean Field approximation a third order
phase transition \cite{BinderYoung} has been found between those two
phases \cite{MEPAVI,JSP} separated by the de Almedia-Thouless line
.\cite{AT} See references \onlinecite{JSP} for a description and
\onlinecite{NewmanStein} for a critic of the RSB picture.

Hence, those two competing theories have opposite predictions about
the overall behavior of a spin glass in the presence of a magnetic field.
However, to perform experimental or numerical tests of the previous
analytical predictions has proved very difficult despite the clear
theoretical predictions (phase transition or not!).

A further step in Mean Field computations is to take into account the
effect of fluctuations. This can be done, for instance, using Field
Theoretic methods and has been done in the past. Working with a
projected theory (by taking only the replicon sector which contains
the most divergent terms of the initial Hamiltonian) no fixed points have
 been found
 in the model.\cite{BRAYROB} It is important to mention
that the existence or not of a transition in magnetic field affects the
existence or not of a phase with replica symmetry breaking (RSB) at
zero magnetic field. In particular the absence of a fixed point for the
Ising spin glass in a magnetic field supports the droplet picture
against the Mean Field one.

A recent and detailed analysis \cite{DeDominicis} reaches the same
conclusions of Ref. \onlinecite{BRAYROB}, and so three possibilities
are opened: {\it i}) no phase transition at all, {\it ii}) first order phase 
transition driven by fluctuations and finally {\it iii}) second order
phase transition dominated by a fixed point outside the accessible
perturbative region (i.e. region of small parameters).  However, a
fourth possibility has been recently opened by Temesv{\'a}ri and De
Dominicis,\cite{TemesvariDeDominicis} who have extended the field
theoretic analysis further. They analyze a field theory in which the
replicon and anomalous sectors are both {\em critical} going beyond
the old analysis where only the replicon sector was taken
critical. The main result of this analysis is the existence of a new
critical point (which appears at eight dimensions) taking the control
of the phase transition at six dimensions. The authors also pointed
out that this new fixed point provides a phase transition which has
different features to that of Mean Field.

On the numerical side the situation is a bit clearer (but not
enough!).  Looking at the off equilibrium numerical simulations, the
difference between the mean overlap and the minimum overlap has been
computed.\cite{PARIRU,MAPAZU,MAPAZU2} In four dimensions and not for
too low temperatures there is a clear difference between these two
measurements, which is a clear signature for Replica Symmetry
Breaking.  Another off equilibrium approach is to compute the
violation of the fluctuation dissipation theorem out of
equilibrium. This has been done using a slightly modified version of
the four dimensional Gaussian spin glass,\cite{PARIRU} and it was
found that the violation is well understood in terms of a non trivial
low temperature phase. In this paper we will follow this approach but
working in three dimensions and simulating the Edwards-Anderson
model. It is important to notice, that the same kind of studies on the
violation of the fluctuation-dissipation out of equilibrium can be
done in real experiments. In fact, in a recent paper, \cite{FDTexp}
the violation of the fluctuation-dissipation relations in an Ising
spin glass (in zero magnetic field) has been reported, and the
experiment can be explained in terms of Replica Symmetry Breaking.

Another numerical method is to use exact ground state techniques in
order to understand the qualitative features of the low temperature
phase. This has been done in three dimensions and a RSB behavior has
been found between zero and a magnetic threshold (for the Gaussian
Ising spin glass in $3d$, this threshold is near 0.65), however this
work cannot exclude completely a droplet behavior.
\cite{Krzakala} A recent study \cite{NEW} points out that there is no
phase transition at all (at finite temperature) but they cannot
exclude a critical field below 0.4.

We can cite different numerical studies working at equilibrium. Those
studies have been mainly done in four dimensions in order to avoid the
proximity to the lower critical dimension (which is between two and
three, at least in the Ising spin glass without magnetic field).
Those studies are not fully conclusive, but the existence of a finite
temperature phase transition emerge as the most likely explanation of
the numerical data.\cite{EQUIL,BOOK}

On the experimental side the situation is not clear. There is strong
experimental evidence about an irreversibility line (where the zero
field cooled (ZFC) and the field cooled (FC) magnetization start to be
different), but unfortunately, that line depends on the time, and so,
we have only off equilibrium information about what happens in the
presence of a magnetic field.  In the literature, one can find some
attempts to analyze the scaling behavior of the freezing temperatures
and the conclusion is that no phase transition exists.\cite{Norblad}

However, recent experimental studies based in the
fluctuation-dissipation relations point out the existence of a phase
transition in the presence of magnetic field.\cite{Orbach}  Moreover, a phase
transition has been reported in a Heisenberg spin glass (AuFe) in three
dimensions in the presence of a magnetic field against the droplet
prediction (we recall that the droplet model predicts no phase
transition independently of the number of components of the
spin).\cite{Campbell}

We will study the three dimensional Ising spin glass using an
off-equilibrium approach based on the computation of the
fluctuation-dissipation relations. This method has provided and
important tool to investigate the low temperature properties of disordered
systems (and it has been  very useful in the  study of non
disordered systems such as glasses).\cite{FRARIE,FDT,FDT2,GLASS}

\section{Theoretical basis}

We have focused this paper on the study of the fluctuation dissipation
relations in the off-equilibrium regime. To do this we need to define
the spin-spin autocorrelation function and the response of the
magnetization to a small change of the magnetic field of the system.

In order to make the paper self-contained and to fix the notation, we shall 
recall some important 
results about the off equilibrium fluctuation-dissipation relations.
We have simulated the binary Ising spin glass in three dimensions on
a cubic lattice of volume $V=L^3$ with helical  boundary
conditions.  
The Hamiltonian of the system is given by
\begin{equation}
{\cal H}=-\sum_{<ij>}\sigma_i J_{ij}\sigma_j -h \sum_i\sigma_i \ .
\protect\label{ham}
\end{equation}
By $<ij>$ we denote the sum over nearest neighbor pairs. The $J_{ij}$
are chosen from $\{+1,-1\}$ randomly and  $h$ is
the external magnetic field.  We have studied systems
with magnetic field $h=0.2$ and lattice sizes
$L=20$, $L=30$  and $L=60$ 

We have used the SUE parallel computer. SUE is a dedicated
machine with an overall performance of 0.22 ns per spin flip. See references
\onlinecite{SUE,SUESIN} for a detailed description of this computer.

Given a quantity $A(t)$ that depends on the local variables of our
original Hamiltonian ($\cal H$), we can define the following
autocorrelation function

\begin{equation}
C(t_1,t_2) \equiv \langle A(t_1) A(t_2) \rangle \ ,
\label{auto}
\end{equation}
and the response function
\begin{equation}
R(t_1,t_2) \equiv \left. \frac{\delta \langle A(t_1) \rangle}{\delta
\hp(t_2)}\right|_{\hp=0} \ ,
\label{res}
\end{equation}
where we have assumed that the original Hamiltonian has been perturbed
to 
\begin{equation}
{\cal H}^\prime= {\cal H} + \int \hp(t) A(t)\,d t \ .
\end{equation}
The brackets $\langle \cdots \rangle$ in eq. (\ref{auto}) and
eq.~(\ref{res}) imply here a double average, one over the dynamical
process and one over the disorder.

As usual one could choose $A(t)=\sigma_i(t)$
and the response function should be
\begin{equation}
R(t_1,t_2) \equiv \left. \frac{\delta m(t_1)}{\delta
\hp(t_2)}\right|_{\hp=0}=\left. \frac{\delta \langle
\sigma_i(t_1)\rangle}{\delta \hp(t_2)}\right|_{\hp=0} \ ,
\label{res1}
\end{equation}
where $m(t)=\langle \sigma_i(t)\rangle$ 

However, to improve the
signal of the autocorrelation we have used in the present paper:
\begin{equation}
C(t_1,t_2) \equiv \frac{1}{V} \sum_{i=1}^V\langle \sigma_i(t_1) 
\sigma_i(t_2) \rangle \ ,
\label{auto1}
\end{equation}
and 
$m(t)=\frac{1}{V} \sum_i \langle \sigma_i(t) \rangle$.  
We remark that we are interested in global
fluctuation dissipation relations. Recently work has been done in
local microscopic fluctuation dissipation relations,\cite{LOCAL} but
we will not study them in this paper.

In the dynamical framework, assuming time translational invariance, it
is possible to derive the fluctuation-dissipation theorem (FDT), that reads
\begin{equation}
R(t_1,t_2)=\beta \theta(t_1-t_2) \frac{\partial
C(t_1,t_2)}{\partial t_2} \ ,
\protect\label{FDT}
\end{equation}
where the inverse temperature is $\beta=1/T$.

The fluctuation-dissipation theorem holds in the equilibrium regime,
but in the early times of the dynamics we expect a breakdown of its
validity.  Mean Field studies \cite{CUKU} suggest the following
modification of the FDT:
\begin{equation}
R(t_1,t_2)=\beta X(t_1,t_2) \theta(t_1-t_2) \frac{\partial
C(t_1,t_2)}{\partial t_2} \ .
\end{equation}
It has also been suggested in \cite{CUKU,FM,BCKP} that the function
$X(t,t^\prime)$ is only a function of the autocorrelation:
$X(t,t^\prime)=X(C(t,t^\prime))$.  We can then write the following
generalization of FDT, which should hold in early times of the
dynamics, the off-equilibrium fluctuation-dissipation relation (OFDR),
that reads
\begin{equation}
R(t_1,t_2)=\beta X(C(t_1,t_2)) \theta(t_1-t_2) \frac{\partial
C(t_1,t_2)}{\partial t_2}\ .
\protect\label{OFDR}
\end{equation}
We can use the previous formula, eq. (\ref{OFDR}), to relate the
observable quantities defined in eq. (\ref{auto}) and eq.~(\ref{res}).
Using the functional Taylor expansion we can write
\begin{equation}
\begin{array}{l}
m[h+\hp](t) = m[h](t) \\\\
+\displaystyle\int_{-\infty}^t d
t^\prime ~ \left.\frac{\delta m[h'](t)}{\delta h'(t^\prime)}
\right|_{h'(t)=h(t)} \hp(t^\prime) + {\mathrm O}(\hp^2) \,,
\end{array}
\end{equation}
and so,
\begin{equation}
\Delta m[h,\hp](t) = \int_{-\infty}^t d t^\prime ~ R(t,t^\prime)
\hp(t^\prime) + {\mathrm O}(\hp^2) \ .
\protect\label{LR}
\end{equation}
where we have defined $\Delta m[h,\hp](t) \equiv m[h+\hp](t)-m[h](t)$.
Eq. (\ref{LR}) is just the linear-response theorem neglecting
higher orders in $\hp$.
By applying the OFDR we obtain the dependence of the
magnetization with time in a generic time-dependent magnetic field
(with a small strength), $\hp(t)$,
\begin{equation}
\Delta m[h,\hp](t) \simeq \beta \int_{-\infty}^t d t^\prime ~
X[C(t,t^\prime)] \frac{\partial C(t,t^\prime)}{\partial t^\prime}
\hp(t^\prime) \ .
\end{equation}

Next we let the system evolves
with the unperturbed Hamiltonian of eq.~(\ref{ham}) from $t=0$ to
$t=\tw$, the so called waiting time, 
and then we turn on the perturbing magnetic field $\hp$ (hence, the system 
feels a magnetic field $h+\hp$). 
Finally, with this choice of the magnetic
field, we can write (ignoring in our notation the fact that
$\Delta m$ also depends on $\tw$)
\begin{equation}
\Delta m[h,\hp](t) \simeq \hp \beta \int_{\tw}^t d t^\prime ~
X[C(t,t^\prime)] \frac{\partial C(t,t^\prime)}{\partial t^\prime} \ ,
\protect\label{mag_1}
\end{equation}
and by performing the change of variables $u=C(t,t^\prime)$, eq.
(\ref{mag_1}) reads
\begin{equation}
\Delta m[h,\hp](t) \simeq \hp \beta \int_{C(t,\tw)}^1 d u ~
X[u] \ ,
\protect\label{mag_2}
\end{equation}
where we have used the fact that $C(t,t) \equiv 1$ (always true for
Ising spins).  In the equilibrium regime ($X=1$ as the fluctuation-dissipation
theorem  holds) we must obtain
\begin{equation}
\Delta m[h,\hp](t) \simeq \hp \beta (1 - C(t,\tw)) \ ,
\protect\label{mag_fdt}
\end{equation}
i.e. $\Delta m[h,\hp](t)\, T/\hp$ is a linear function of
$C(t,\tw)$ with slope $-1$.  We remark that we can use this formula to
obtain $q_{\mathrm{max}}$ as the point where the curve $\Delta m[h,\hp](t)$
against $C(t,\tw)$ leaves the line with slope $-\beta \hp$.

In the limit $t, \tw \to \infty$ with $C(t,\tw) = q$, one has that
$X(C) \to x(q)$, where $x(q)$ is given by
\begin{equation}
x(q)=\int_{q_{\mathrm{min}}}^q ~d q^\prime ~P(q^\prime)\ , 
\protect\label{x_q}
\end{equation}
where $P(q)$ is the equilibrium probability distribution of the
overlap. Obviously $x(q)$ is equal to 1 for all $q >
q_{\mathrm{max}}$, and we recover FDT for $C(t,\tw) >
q_{\mathrm{max}}$.  This link between the dynamical function $X(C)$
and the static one $x(q)$ has been already verified for finite
dimensional spin glasses.\cite{FDT} The link has been analytically
proved for systems with the property of stochastic stability.\cite{FRANZ}

For future convenience, we define 
\begin{equation}
S(C)\equiv \int_C^1 d q ~x(q)\ ,
\protect\label{s_c}
\end{equation}
or equivalently
\begin{equation}
P(q) =-\left.\frac{d^2 S(C)}{d^2 C}\right|_{C=q} \ .
\label{pq}
\end{equation}  
In the limit where $X \to x$ we can write eq.~(\ref{mag_2}) as
\begin{equation}
\frac{\Delta m[\hp](t)\;T}{\hp} \simeq S(C(t,\tw)) \ .
\protect\label{final}
\end{equation}

Looking at the relation between the correlation function and the
integrated response function for large $\tw$ we can thus obtain
$q_{\mathrm{max}}$, the maximum overlap with non-zero $P(q)$, as the point
where the function $S(C)$ becomes different from the function $1-C$.

From the function $S(C)$ we can get information on the overlap
distribution function $P(q)$, through eq.~(\ref{pq}).  Let us recall
which is the prediction for the $S(C)$ assuming the validity of each one of
the competing theories described in the introduction.  The droplet
model predicts $P(q)=\delta(q-{\hat q})$ and consequently
\begin{equation}
S(C) = \left\{
\begin{array}{cl}
1 - {\hat q} & {\mathrm{for}} \;\; C \le {\hat q} \ ,\\
1 - C & {\mathrm{for}} \;\; C > {\hat q} \ .
\end{array}
\right.
\protect\label{droplet}
\end{equation}

In models with only one state, as the droplet model predicts for this
model, the equilibrium time is finite irrespective of the value of the
volume of the system, hence, we can always thermalize any volume, and
so the asymptotic behavior, for waiting times larger than the
equilibration time, is composed only for the straight line $1-C$. There
is no horizontal part.

  On the other hand the MF like prediction for the overlap
distribution\cite{MEPAVI} 
$P(q) = (1-x_\mathrm{M}) \delta(q-q_{\mathrm{max}}) + x_\mathrm{M}
\delta(q-q_{\mathrm{min}}) + \tilde{p}(q)$ (where the support of
$\tilde{p}(q)$ belongs to the interval
$[q_{\mathrm{min}},q_{\mathrm{max}}]$, $q_{\mathrm{min}} \propto
h^{4/3}$ and $q_{\mathrm{max}}$ mainly depends on the temperature),
implies that
\begin{equation}
S(C) = \left\{
\begin{array}{cl}
S(0) & {\mathrm{for}} \;\; C \le q_{\mathrm{min}} \ ,\\
\tilde{s}(C) & {\mathrm{for}} \;\; q_{\mathrm{min}} < C \le q_{\mathrm{max}} \ ,\\
1 - C & {\mathrm{for}} \;\; C > q_{\mathrm{max}} \ ,
\end{array}
\right.
\protect\label{rsb}
\end{equation}
where $\tilde{s}(C)$ is a quite smooth and monotonically decreasing
function such that
\begin{equation}
\tilde{p}(q) = -\left.\frac{d^2\tilde{s}(C)}{d C^2}\right|_{C=q}\ .
\end{equation}

To finish this section we will recall an approximate scaling property
of the probability distribution of the overlap that was introduced by
Parisi and Thouless (hereafter PaT).\cite{ParisiThouless}
In particular in Mean field the PaT hypothesis implies 

\begin{equation}
S(C) = \left\{
\begin{array}{cl}
1-C & {\mathrm{for}} \;\; C \ge q_{\mathrm{max}} \ ,\\
T \sqrt{1-C} & {\mathrm{for}} \;\; q_\mathrm{min} \le C 
\le q_{\mathrm{max}} \ .\\
\end{array}
\right.
\end{equation}

The result for $C \ge q_{\mathrm{max}}$ is general (and true for
finite dimension) and for $q_\mathrm{min} \le C \le q_{\mathrm{max}}$
we make the following Ansatz: $S(C)= A T (1-C)^B$ (in Mean Field $A=1$
and $B=1/2$).  If we substitute this Ansatz in eq.~(\ref{final})
we obtain the following scaling equation
\begin{equation}
\frac{m T}{h} T^{-\phi} =f\left( (1-C) T^{-\phi} \right)\ ,
\end{equation}
where $f$ is a scaling function and $\phi=1/(1-B)$ (in Mean
Field $\phi=2$). In order to be consistent 
the scaling function should be  composed by a linear part
($x$) and by a power law part ($A x^B$).

We have only measured the autocorrelation function (see 
eq.~(\ref{auto1})) and the response function (see eq.~(\ref{res1})).

\section{Numerical results}

\subsection{On the critical temperature}

Assuming the existence of a phase transition in magnetic field, we can
estimate the shift of the critical temperature when a small magnetic
field is turned on using the Mean Field approximation. The main formula
is \cite{BinderYoung}
\begin{equation}
\frac{\Tc(h)-\Tc(0)}{\Tc(0)}=\left( \frac{3}{4} \right)^{1/3} 
\left(\frac{h}{J} \right)^{2/3} \,,
\label{AT_MF}
\end{equation}
where $J$ is defined (in Mean Field) as ${\overline {J_{ij}^2}}=J^2/N$,
$N$ being the volume of the system (or the coordination number in the
Mean Field approximation), $J_{ij}$ being the random couplings between the
spins, $\Tc(0)$ being the critical temperature in absence of magnetic
field and finally $\Tc(h)$ being the critical temperature in the presence of
a magnetic field $h$. That is the formula that fixes the AT
instability in infinite dimensional spin glasses.

We can modify that formula (eq. (\ref{AT_MF})) for a finite coordination
number. Let $z$ be the coordination number of our
lattice. We recall that our $J_{ij}$ have unit variance and so
$J=\sqrt{z}$. Hence we can write
\begin{equation}
\frac{\Tc(h)-\Tc(0)}{\Tc(0)}=\left( \frac{3}{4} \right)^{1/3} 
\left(\frac{h}{\sqrt{z}} \right)^{2/3} \,,
\label{AT_DF}
\end{equation}
In our case $z=6$, $\Tc(0)\simeq 1.14$, and so
$T(h=0.2) \simeq 0.945$ ($h=0.2$ is the magnetic field simulated in
the present work). Notice that near zero magnetic field the phase
transition line has vertical slope ($dT(h)/dh \simeq 1/h^{1/3}$).

In order to check the existence or not of a phase transition (using
the OFDR as a tool) we have simulated at very high temperature
($T=2.5$) and a lower temperature (which is below our previous 
estimate of the critical one, $T=0.714$).

\subsection{OFDR in the high temperature region}

We have simulated the system at temperature $T=2.5$ in a magnetic
field $h=0.2$ and with perturbing fields $\hp=0.01$ and $\hp=0.03$ (in
order to check linear response) and different waiting times: 409600
and 
819200. The number of samples simulated was about 6400
samples for each waiting time.

\begin{figure}
\includegraphics[width=\columnwidth]{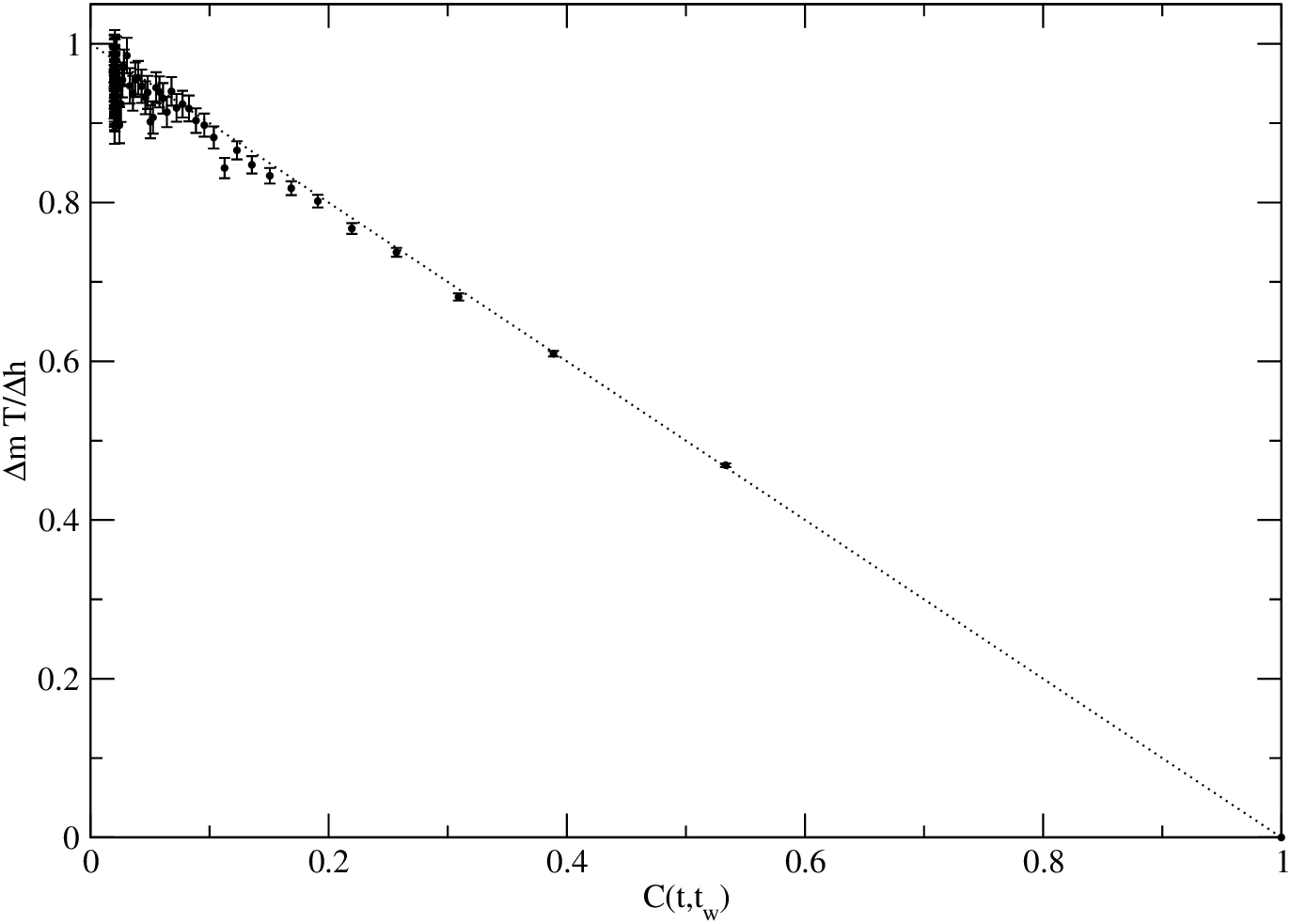}
\caption{Off equilibrium fluctuation-dissipation relations for
$T=2.5$, $h=0.2$ and $L=30$. We have drawn the equilibrium
straight line $1-C$.  We plot $\Delta m T/h$ against $C(t,\tw)$ for
the waiting time $\tw=819200$.}
\label{HIGH_T}
\end{figure}

We show in figure~\ref{HIGH_T} the plot for $\tw=819200$ and $L=30$.

For the largest value of the waiting time simulated all the data stay
on the equilibrium line $1-C$ (i.e. this waiting time is greater than
the equilibration time for this lattice size).

\begin{figure}
\includegraphics[width=\columnwidth]{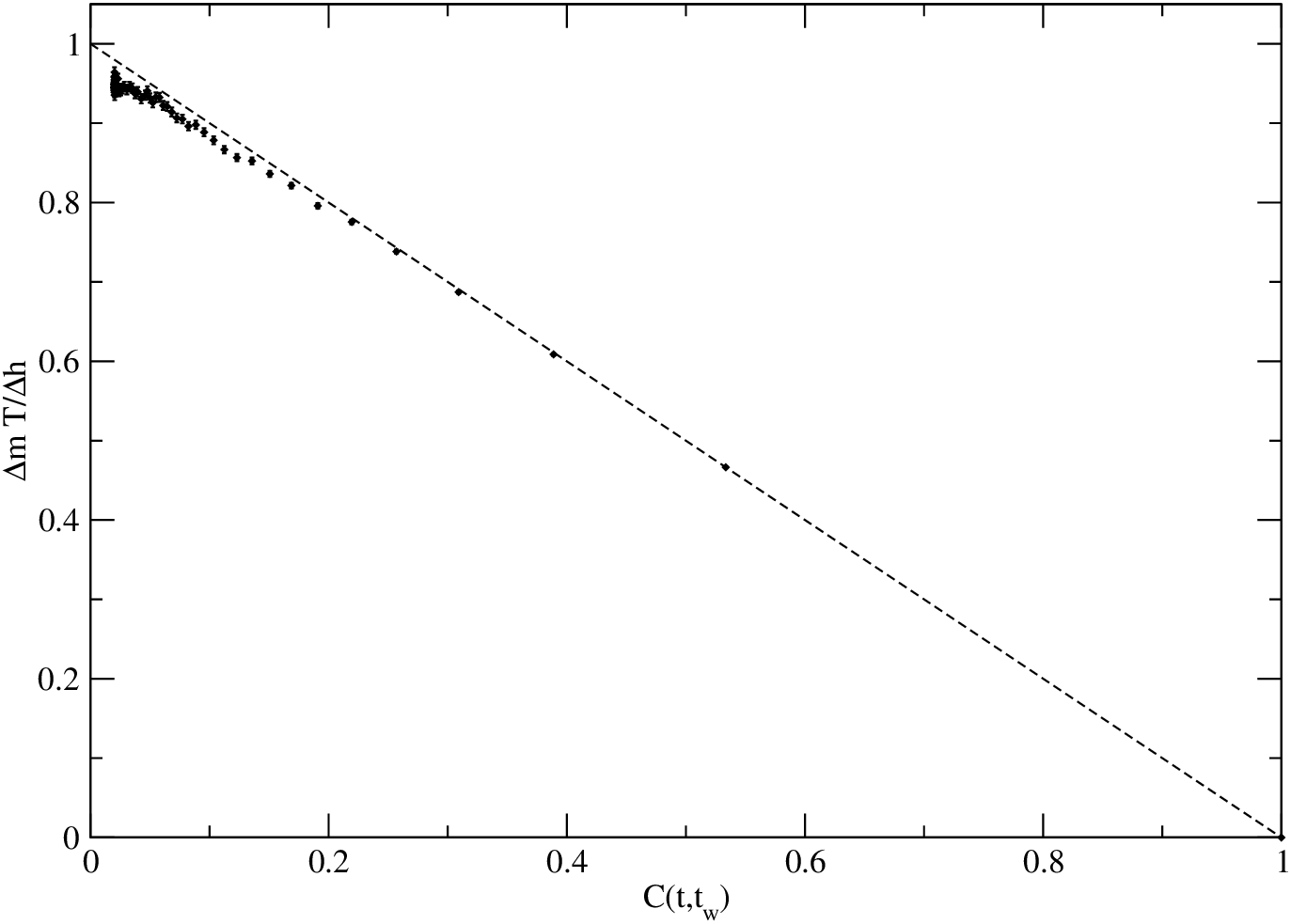}
\caption{As figure \ref{HIGH_T} but with $\tw=409600$.}
\label{HIGH_T_B}
\end{figure}

In the paramagnetic phase, droplet and RSB agree: for a ``finite''
volume and very large times (grater than the equilibration time) all
the points should lie on the straight line $\Delta m(t) T/\hp=1-C$
with $C\in[0,\qEA]$ (equilibrium situation). For intermediate
situations (i.e. not so large waiting times) the curves lie below the
straight line (see figure \ref{HIGH_T_B}) and the final straight line,
is built from below (i.e. curves with lower waiting times lie below
those with higher ones). This behavior is similar to that found in the
two dimensional Ising spin glass in a magnetic field for a finite
temperature (the system is paramagnetic).\cite{BarratBerthier} See
also figure 5 of reference \cite{PARIRU} for an example of a FDT plot
in a paramagnetic phase in the four dimensional Ising spin glass.

\subsection{OFDR in the low temperature region}

The situation at a lower temperature is dramatically different.

We start showing the numerical results for one of the lowest temperature
simulated, $T=0.714$. All the simulations reported in this subsection have 
been done at $h=0.2$ as the external magnetic field.

\begin{figure}
\includegraphics[width=\columnwidth]{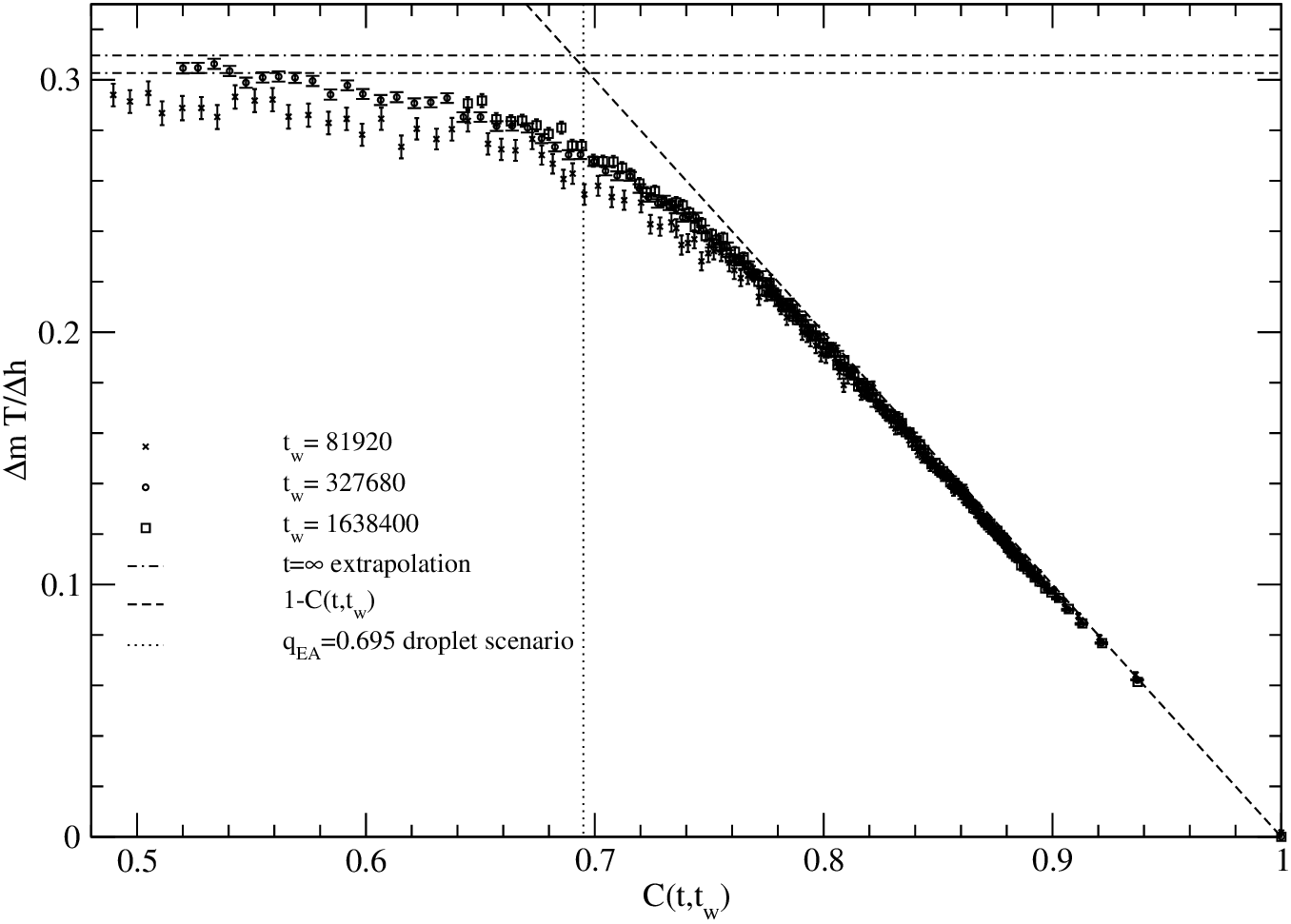}
\caption{Off equilibrium fluctuation-dissipation relations for
$T=0.714$, $L=30$ and $h=0.2$. We have marked the equilibrium straight
line $1-C$. We plot $\Delta m T/h$ against $C(t,\tw)$ for three
different waiting times. We have also plotted the error band for the
asymptotic value of $\Delta m T/h$.}
\label{LOW_T}
\end{figure}

In figure \ref{LOW_T} we show $ \Delta m T/\hp$ against $C(t,\tw)$
for different waiting times $\tw$ and perturbing magnetic field
$\hp=0.03$ for the $L=30$ lattice.

The first check we have performed is to control that we are in the
linear response regime. To do this we have computed the OFDR for
different perturbing magnetic fields $\hp=0.01$ and $0.03$. We have
found that the results are independent of these two values of
$\hp$. In the following we show the results obtained with $\hp=0.03$.

\begin{figure}
\includegraphics[width=\columnwidth]{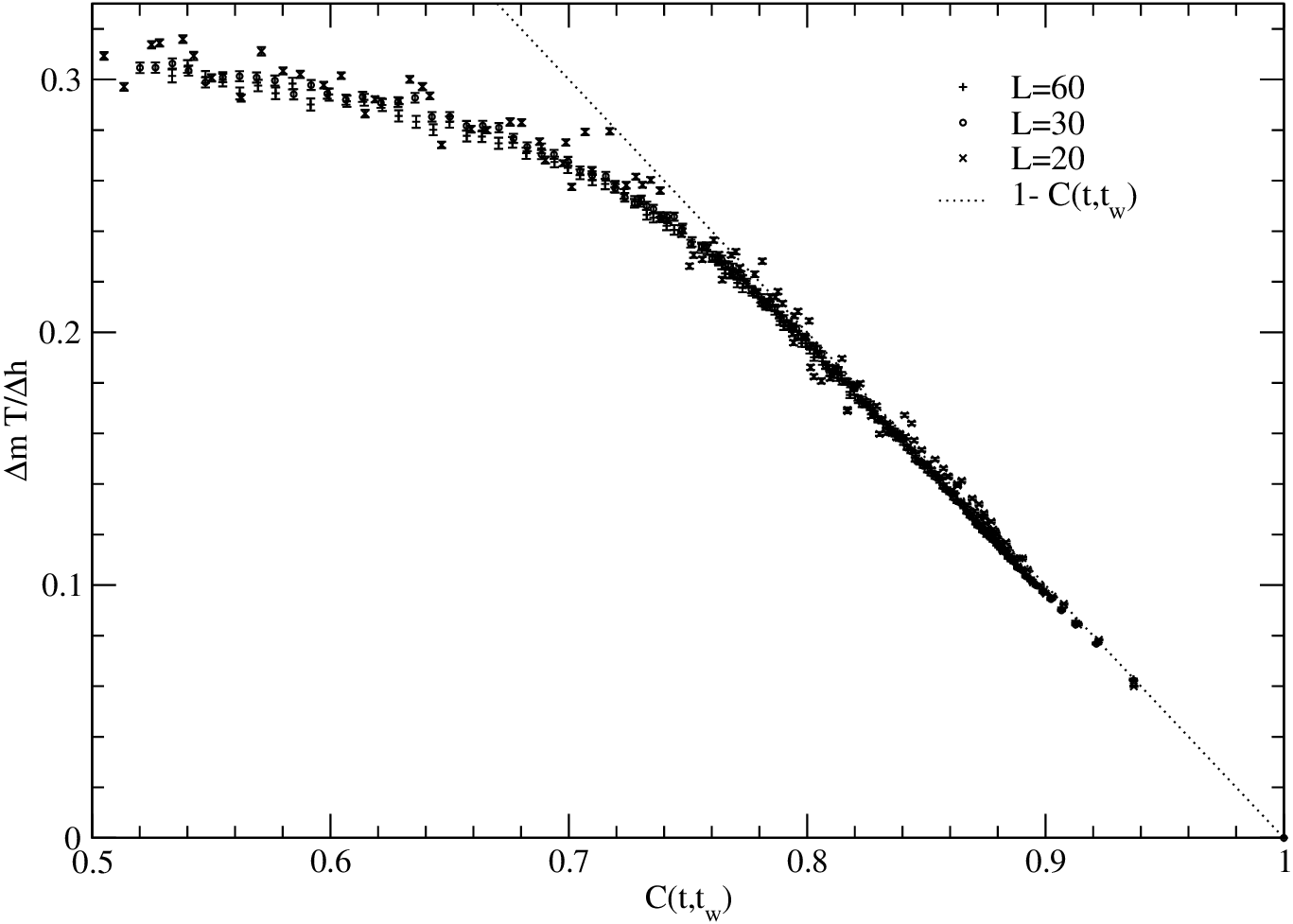}
\caption{Off equilibrium fluctuation-dissipation relations for
$T=0.714$ and three different lattice sizes $L=60$, $L=30$,
$L=20$. The perturbing field is $0.03$ for $L=30$, $60$ lattices and $0.06$ for
$L=20$ lattice. We show the data for one of the larger waiting time
simulated $\tw=327680$.}
\label{20_LOW_T}
\end{figure}

The second check has been to verify that our results are lattice size
independent. In figure \ref{20_LOW_T} we can see the results for
 $L=30$ and  $L=60$, with perturbing field $\Delta h =0.03$. 
For $L=20$, the perturbing field $\Delta h =0.03$ has proved too noisy, 
and we have used $\Delta h =0.06$.
It can
be seen that the behavior of $L=30$ is asymptotic in this kind of
simulations (i.e. for the time scales that we have simulated): the
$L=20$ points are still a bit noisy but $L=30$ and $L=60$ coincide.  Using this
information we will focus on the $L=30$ lattice in the rest of the
paper. We can state that we have simulated 512 samples for $\tw=81920$,
416 samples for $\tw=163840$ and 3232 for $\tw=327680$ and $1638400$ in the $L=30$
lattice. In addition $6200$ samples in the $L=20$ lattice and $412$ in the
largest lattice that we have simulated ($L=60$).

The third goal is to study the dependence on the waiting times of the
OFDR curves. We have simulated $\tw=81920$, $163840$, $327680$ and
$1638400$  and
from figure \ref{LOW_T} it is possible to see that the curves rise
as the waiting time is larger.  Moreover, the
curves for the larger waiting times are just compatible within our error
bars. In this sense we are confident that the curve corresponding to
$\tw=1638400$ represent very well the overall behavior of the system
(very large volumes and times or equivalently infinite volume and
waiting times).  This behavior is very important
because our final curve (i.e. $\tw=1638400$) has a clear curvature
which should be absent if the droplet model holds (see eqs.~(\ref{droplet}) 
for droplet and (\ref{rsb}) for RSB predictions).
     
\begin{figure}
\includegraphics[width=\columnwidth]{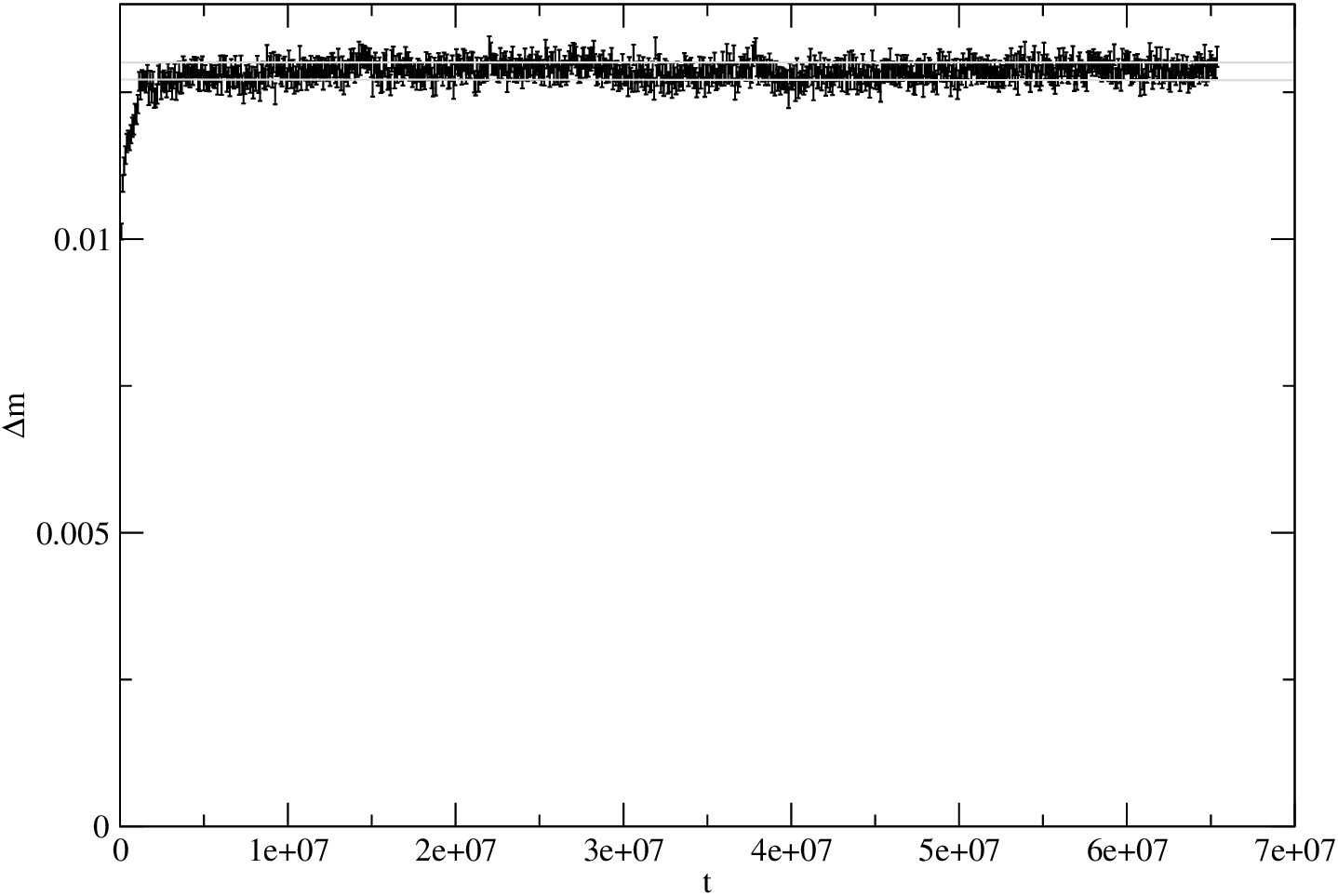}
\caption{$\Delta m(t)$ against $t$ for $\tw=327680$ and $L=30$ in a
simulation longer in time than the ones used in the figures of FDT, but
with much less samples $1152$. The fit to equilibrium plateau is also
shown (we show again the error band for the mean value).}
\label{MAG_LOW_T}
\end{figure}

We have plotted in figure \ref{LOW_T} two additional straight lines.
The first line, horizontal, corresponds to the asymptotic value
of $\Delta m(t) T/\hp$. To obtain this value, we have performed a
simulation reaching times much longer than the ones used in the OFDR
curves, but using a smaller number of samples. In figure
\ref{MAG_LOW_T} we can see the evolution of $m(t)$. The horizontal
lines are the error band related to 
the asymptotic value of $\Delta m(t)=0.1286(15)$. For this value we do
not need to do any kind of extrapolation since we have reached the
thermodynamic value of $\Delta m(t) T/\hp$ in this long simulation
($O(10^8)$ Monte Carlo steps).   

If the droplet model holds, the final (asymptotic) curve should be
composed by the $1-C$ straight line ($C \in[\qEA,1]$) as
explained above.  This implies that we can estimate the ``droplet''
prediction for the order parameter as $\qEA \simeq
0.694(4)$. We have marked this value with a vertical line in figure
\ref{LOW_T}.
 
We would remark at the end of  this section the following points:
\begin{itemize}

\item We have obtained a $\tw$-independent final curve (and
$L$-independent), at least within our statistical precision.  We
believe that this curve represents with high accuracy the behavior for
large volumes and times of an Ising spin-glass at $T=0.714$ and
$h=0.2$. In this scenario we can estimate that $\qEA\simeq 0.76(2)$
(the points in which the points leave the straight line) which differs
from the droplet value $0.694(4)$.

\item We cannot avoid a dependence on the waiting times beyond our
numerical precision, and so we can not exclude completely a droplet
phase with $\qEA=1-\Delta m(\infty) T/\hp\simeq 0.694(4)$.  

\item As we have cited in the section devoted to high temperature, the
final curve is built from below. At low temperature, in the droplet
scenario, we should expect the same behavior as at high temperature
and so the final curve should build from below. The point is whether the
curves for large waiting times stop or not before they reach the
droplet prediction. Our numerical data suggest that the curves stop
before the droplet final curve, and that the asymptotic curve shows the
characteristic curvature of a phase with RSB.

\end{itemize}

\begin{figure}
\includegraphics[width=\columnwidth]{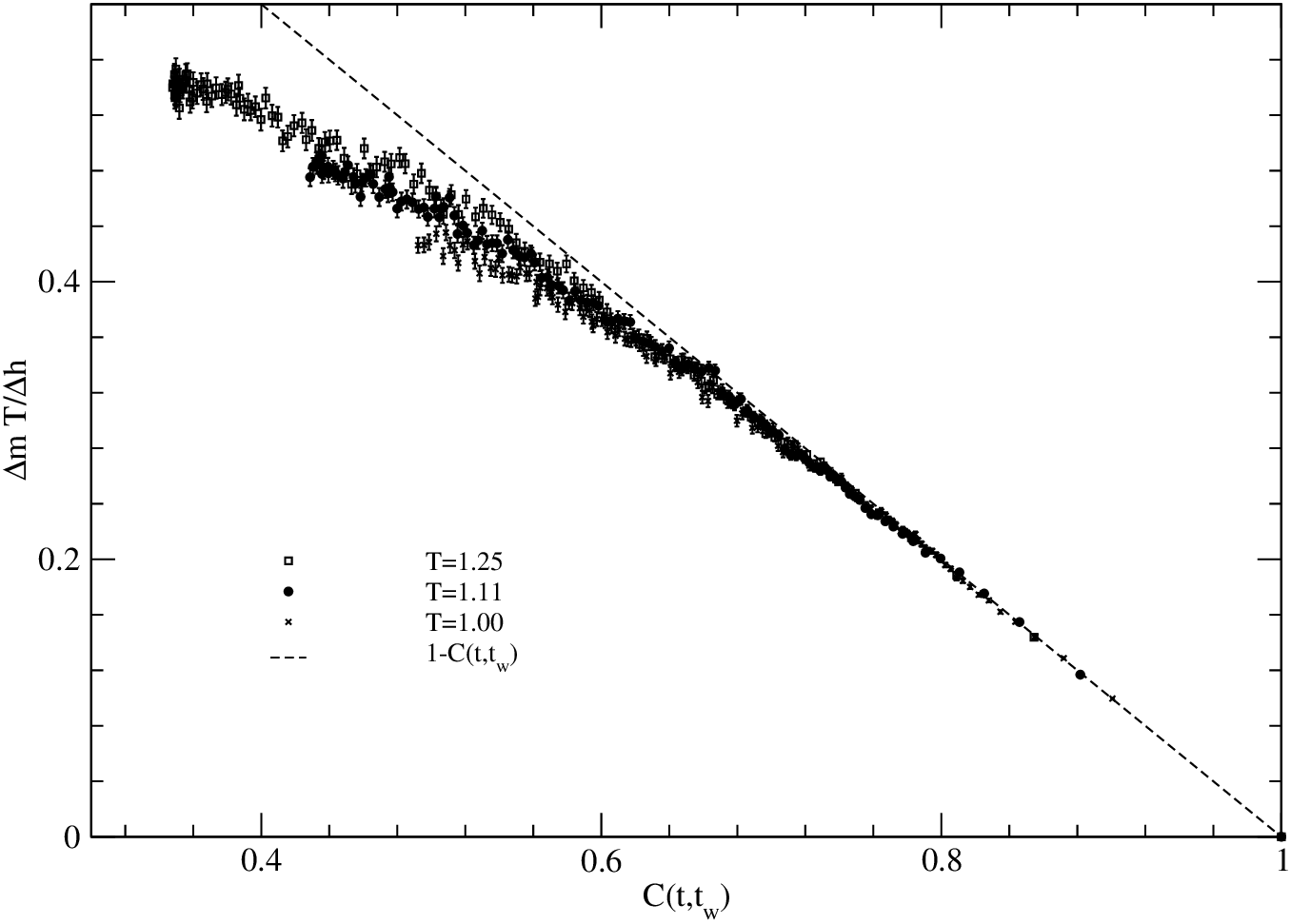}
\caption{We show OFDR for three temperatures, for $\tw=327680$ and $L=30$.}
\label{crossover}
\end{figure}

We end this section showing a figure corresponding to the crossover
region. In the following discussion we will restrict ourselves to a
qualitative level.  In figure \ref{crossover} we have shown the OFDR
for the three values of temperature ($T=1.25$, $1.11$ and $1.0$) and
$t_w=327680$. It is clear that the largest temperatures shows a clear
signature of a paramagnetic phase (i.e. small curvature and almost a
straight line). You can compare these curves with a clear paramagnetic
one, see figure \ref{HIGH_T_B}, which also shows a small curvature at
the end of the curve. We remark that the critical temperature of the
model with no magnetic field is about $1.14$, and so we are
(qualitatively) exploring the region near the vertical of this
point. On the basis of mean field we should expect that the line of
transitions emerges from the point at zero magnetic field with
vertical slope. Hence, this plot and a transition temperature near
$1.1$ is compatible with this scenario. The prediction using 
eq.~(\ref{AT_DF}) was $0.95$. Obviously, in order to see a clear
paramagnetic curve with no violations, we refer to figure
\ref{HIGH_T}. We remark that this methods based in the violation of
fluctuation-dissipation is not so powerful to determine with precision
transition points; is a good method to decide if one point $(T,h)$
(well inside the phase in order to avoid the crossover region) behaves
in a way or not.

\subsection{OFDR in the high magnetic field region}

In this section we study the properties of OFDR at a fixed low
temperature when the magnetic field  grows. The goal of this section is
to find when the behavior of the OFDR relations change from a non
trivial one (as found for $T=0.714$ and $h=0.2$) to a trivial one
(droplet) as far the magnetic field becomes larger. Notice that this
temperature ($T=0.714$) is far away of the critical temperaure of the
model with no magnetic field ($T=1.138$), avoiding so, crossover
effects between the phase transition at zero field.

\begin{figure}
\includegraphics[width=\columnwidth]{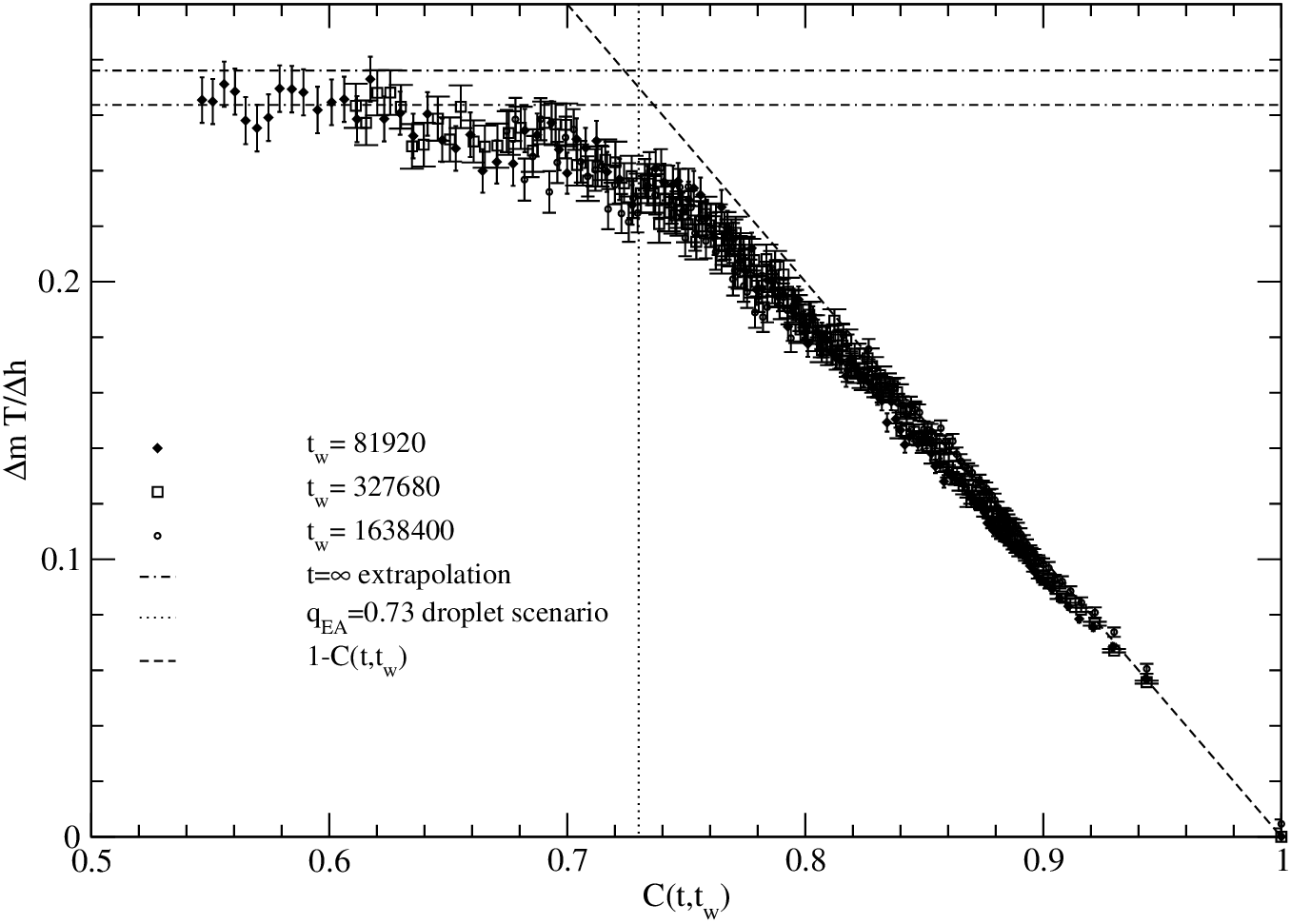}
\caption{Off equilibrium fluctuation-dissipation relations for
$T=0.714$, $L=30$ and $h=0.4$. We have marked the equilibrium straight
line $1-C$. We plot $\Delta m T/h$ against $C(t,\tw)$ for three
different waiting times. We have also plotted the error band for the
asymptotic value of $\Delta m T/h$.}
\label{LOW_T_H04}
\end{figure}

\begin{figure}
\includegraphics[width=\columnwidth]{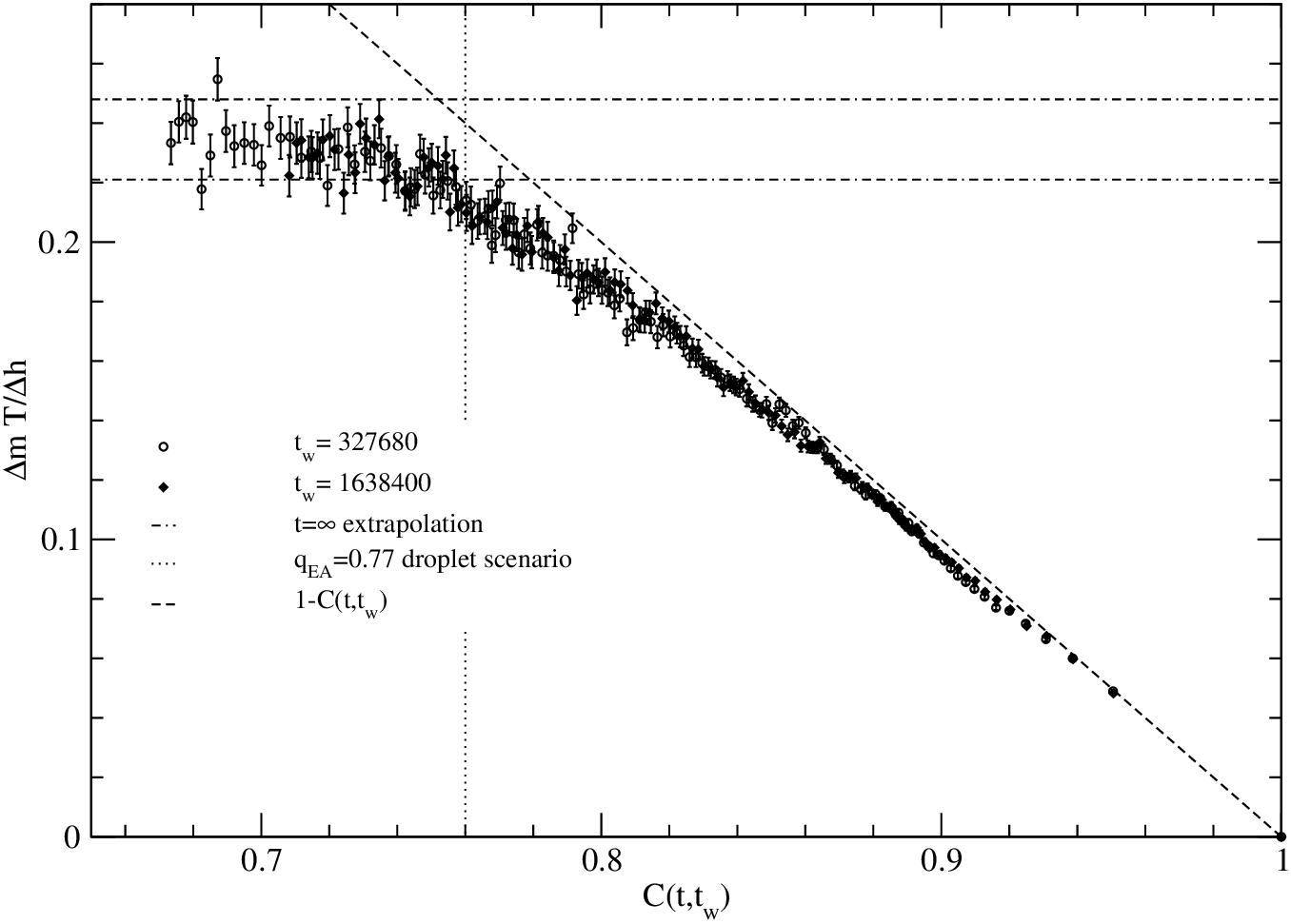}
\caption{Off equilibrium fluctuation-dissipation relations for
$T=0.714$, $L=30$ and $h=0.6$. We have marked the equilibrium straight
line $1-C$. We plot $\Delta m T/h$ against $C(t,\tw)$ for two
different waiting times. We have also plotted the error band for the
asymptotic value of $\Delta m T/h$.}
\label{LOW_T_H06}
\end{figure}

We show in figures \ref{LOW_T_H04} and \ref{LOW_T_H06} the
results obtained at $T=0.714$ and $h=0.4$ and $h=0.6$ respectively.

We start discussing the $h=0.4$ plot (figure \ref{LOW_T_H04}). If we
compute, as above, $\qEA$ as the minimum value of the
correlation (starting from $C=1$) for which the points do not lie
(using one standard deviation as critera) on the straight line, we
obtain that this value of the field 
is still not statistically compatible with the
droplet value. If we relax the one standard deviation criteria to two or three
standard deviation, the behavior can be described as droplet.
Obviously for larger magnetic field we can show analytically that the
behavior is droplet (the magnetic contribution in the Hamiltonian
becomes dominant and we can drop the spin glass term).

For a larger magnetic field, $h=0.6$, the situation is clearer (figure
\ref{LOW_T_H06}). Practically all the points are on the straight
line (slightly below but always at a distance less than a one-two  standard
deviations). We have a horizontal part (the system has reached
its asymptotic magnetization) since our largest waiting time is still
smaller than the equilibration time for this magnetic field (in the
droplet, this time is finite).

The conclusion of this section is that $h=0.4$ the situation is still
not clear but $h=0.6$ is droplet.  We have observed a clear change in
the behavior of OFDR in the region $h\simeq 0.4-0.6$. Assuming a phase
transition (between RSB and droplet) this implies that $\hc(T=0.714)
\sim 0.6$.

This figure compares very well with that obtained at zero temperature
in a related model (which uses Gaussian coupling). It has been
obtained a critical magnetic field at zero temperature of 0.6
\cite{Krzakala} or 0.4 if we use reference \onlinecite{NEW}).

\subsection{Scaling properties of OFDR in the low temperature region}

In this section we will study the scaling properties of the $\tw$ and
$L$-independent fluctuation dissipation curves obtained for different
temperatures. The main goal of this section is to study the degree of
accuracy of the approximate PaT Ansatz in the three dimensional Ising
spin glass in a magnetic field. That Ansatz has been found very
adequate to describe the low temperature fluctuation-dissipation
curves both in three and four dimensions in the presence of a magnetic
field. Moreover this scaling Ansatz has been checked in experiments
finding that it describes very well the experimental data. The Ansatz
has also been studied in the two dimensional spin glass (no phase
transition) and it has been found that the curves computed for
different temperatures at the same $\tw$ ($\tw=10^4$) for a large
lattice ($V=400^2$ ) also follow this Ansatz (see figure 1-a of
reference \onlinecite{BarratBerthier}). Notice that this waiting
time is not an asymptotic one (see fig 1-b of reference 
\onlinecite{BarratBerthier}).

\begin{figure}
\includegraphics[width=\columnwidth]{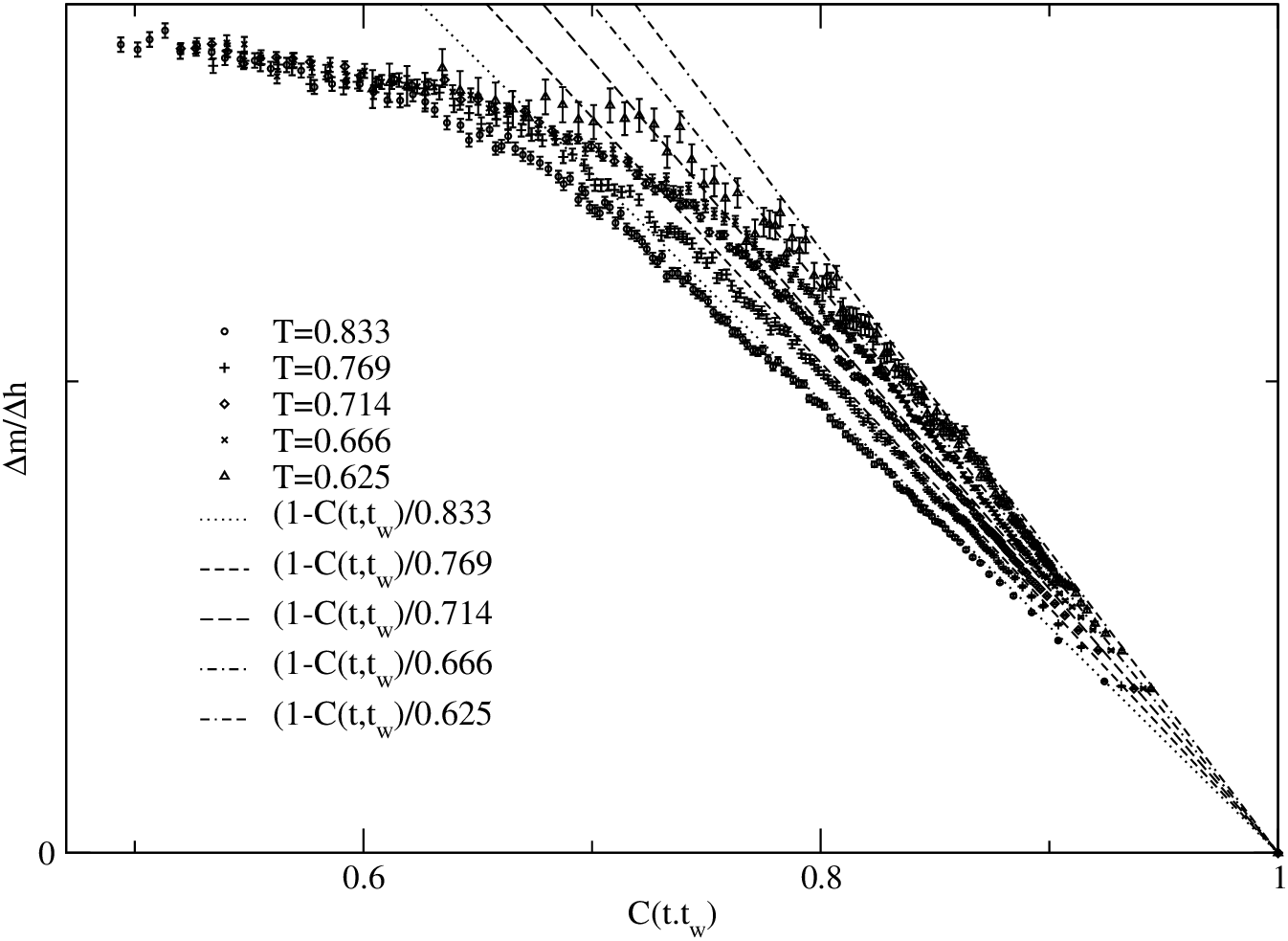}
\caption{Off equilibrium fluctuation-dissipation relations for five
different temperatures. We have marked the equilibrium straight
lines. Notice that we plot $\Delta m/h$ instead $\Delta m T/h$ as in
previous plots, hence the slopes of the equilibrium lines are
$1/T$. We show only the data computed with the largest waiting time
$\tw=327680$.}
\label{ALL_T}
\end{figure}

\begin{figure}
\includegraphics[width=\columnwidth]{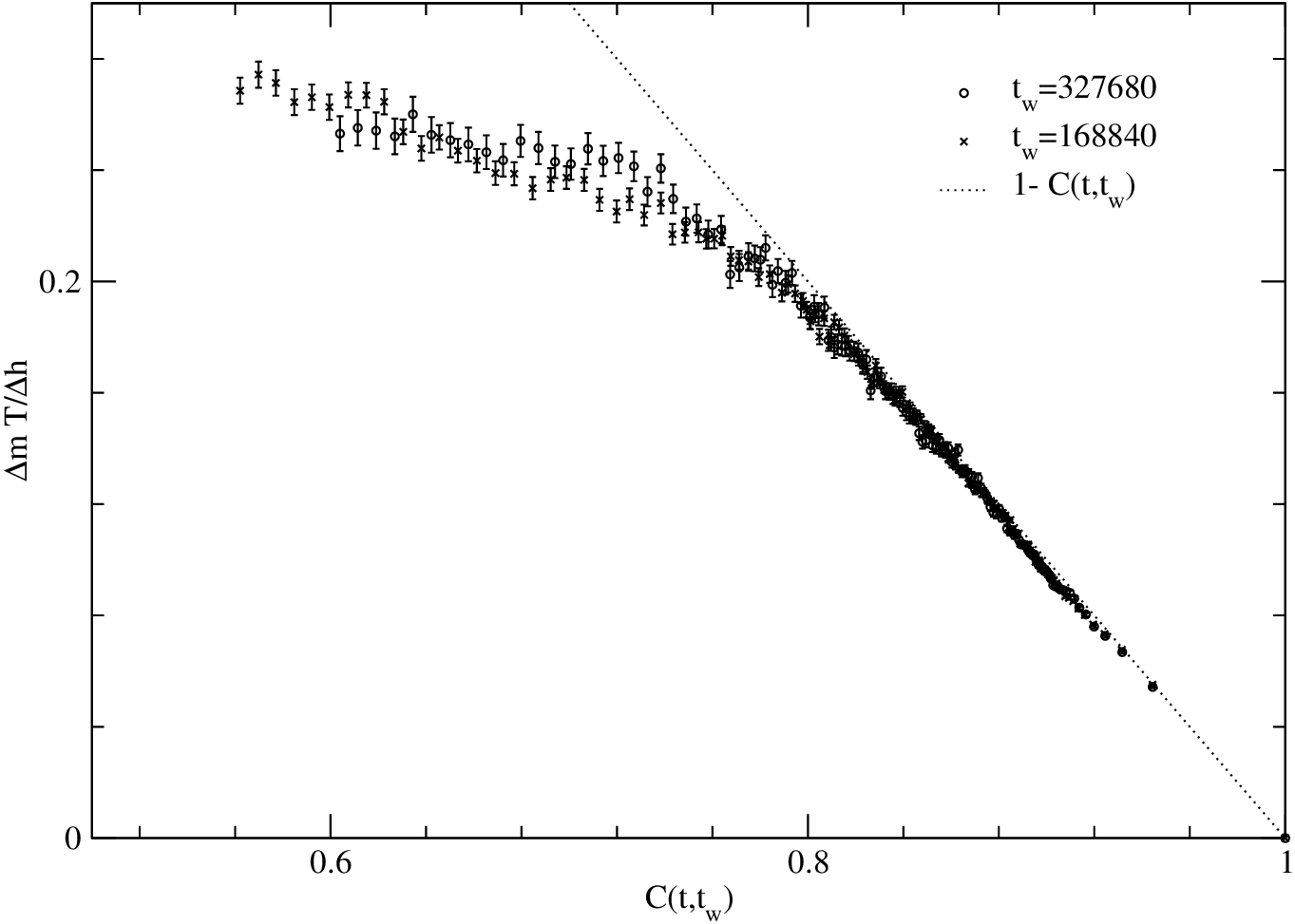}
\caption{Off equilibrium fluctuation-dissipation relations for
$T=0.625$ (the coldest temperature we have simulated) for $L=30$,
$h=0.2$ and $\Delta h=0.03$. We show the data for the largest waiting
time simulated $\tw=327680$ and for $\tw=162840$. The date seem to be
asymptotic in the statistical error.}
\label{FDT_LOW_T}
\end{figure}

We show in figure \ref{ALL_T} the $\tw$ and $L$-independent
fluctuation dissipation curves obtained at five different
temperatures. In particular, we show in figure \ref{FDT_LOW_T} the
fluctuation-dissipation plot for the lowest temperature we have
simulated, $T=0.625$, and for two different and large waiting 
times ($\tw=162840$
and $\tw=327680$). It is clear that these two waiting times are
asymptotic (within error bars) and so we are again confident
in that our curves plotted in figure \ref{ALL_T} are asymptotic.

We have deleted the temperature factor in the ordinate axis and so
each FDT straight line has slope $1/T$. We can see that when the data
leave the pseudo-equilibrium region (i.e. the straight line), they go
to the same curve (independently of their temperatures). This is a
strong signature that the PaT Ansatz works.

\begin{figure}
\includegraphics[width=\columnwidth]{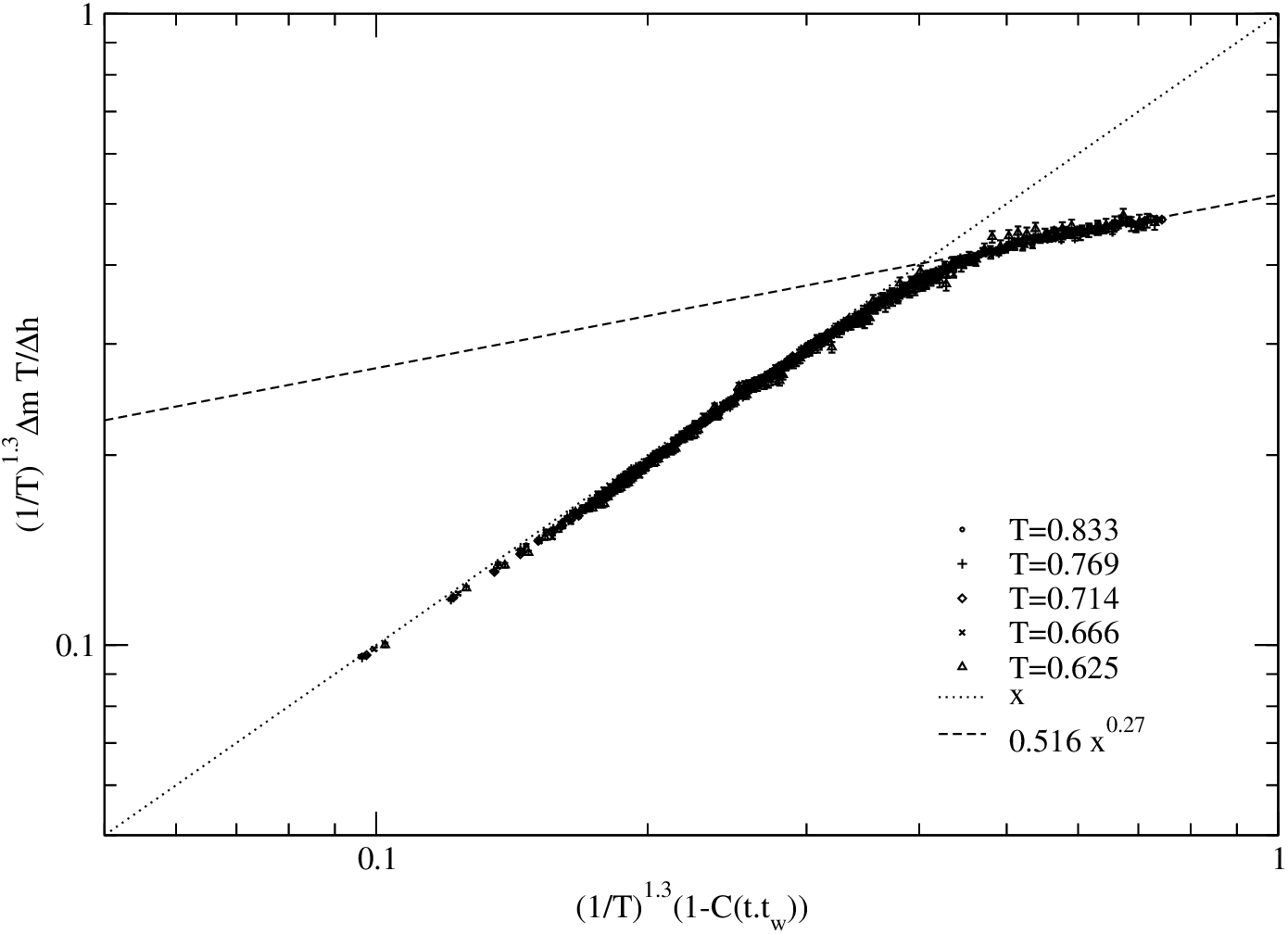}
\caption{Scaling plot of the off equilibrium fluctuation-dissipation
relations for four different temperatures in the low temperature region.}
\label{ALL_T_SCALING}
\end{figure}

We have tried in figure \ref{ALL_T_SCALING} this kind of Ansatz and
the scaling is very good and so we are confident that the PaT scaling
describes with great accuracy the behavior of the fluctuation
dissipation curves in a large temperature window.

Notice that the PaT scaling works for our $L-$ and $\tw-$independent
curves. We have found a good scaling for values of $\phi\in
(1.2,1.4)$. In figure \ref{ALL_T_SCALING} we use central value for
$\phi=1.3$. Two clear and distinctive regimes can be seen in that
figure . The first one correspond to the quasi-equilibrium regime: in
that part of the figure the behavior is linear and so it matches with
the quasi-equilibrium regime $\Delta m T /h = 1-C$. The second one
corresponds to the aging regime: that part of the plot can be
parametrized with a power law with the $B$ exponent introduced above
in the paper. We have obtained $B=0.27(3)$ which provides
$\phi=1.37(6)$, which is a compatible value with the $\phi$ value used
in the scaling plot (this is a check of consistency of the scaling
law!). For completeness we report the value of $A$ (we remark that the
aging region follows a law $A x^B$, where $x$ is the scaling variable
$T^{-\phi} (1-C)$): $A=0.52(1)$.

We can compare the values obtained for $A$ and $B$ with previous
results published in the literature. In the $3d$ Ising spin glass with
no magnetic field $A=0.7$ and $B=0.41$.\cite{PAR_3D} For the $4d$
Ising spin glass again with no magnetic field $A=0.52$ and $B=0.41$.
\cite{FDT}

We can see that in absence of magnetic field the $B$ value is close to the
Mean Field value (0.5) whereas the magnetic field value in three
dimensions is clearly far from the MF value.

Following reference \onlinecite{BarratBerthier} this kind of scaling it is
not enough to detect a RSB phase (they found in the two dimensional
Ising model ---with no phase transition at finite temperature--- a PaT
scaling for their OFDR). Nevertheless, in \cite{BarratBerthier} the
PaT scaling only works for points with the same waiting time, instead,
in our plot we have points computed with different waiting times. In
effect, we remark again, our scaling is $\tw$-independent (at least in
our numerical precision) which is a behavior completely different
from the two dimensional spin glass (paramagnetic phase).  For a
paramagnetic phase and very long waiting time (i.e. all the points lie
in the $1-C$ straight line, see for example our figure \ref{HIGH_T},
we have equilibrated the system) the PaT scaling plot should consist
in points over the linear part (quasi-equilibrium regime), and no one
in the power law part (aging regime).

We finally remark that this scaling in addition with the analysis
of the OFDR (see above) provides us with a picture that could be
explained assuming a low temperature phase with RSB.

\section{Discussion and Conclusions}

We have studied how the fluctuation dissipation relations work off
equilibrium in the three dimensional Ising spin glass with a magnetic
field.

We have shown numerical data that has been obtained simulating very
large lattices ($L=20, 30$ and $ 60$) and for extremely large times
for the three dimensional Ising spin glass. In order to achieve these
lattice sizes and time we have used a dedicated machine (SUE).

We have identified with this tool a paramagnetic phase in the
high temperature region (as expected) and a phase where we have found
strong violations of fluctuation-dissipation. We can describe very
well (within our statistical precision) these violations assuming a RSB
scenario, yet we can not exclude completely a droplet scenario.

Moreover we have shown the crossover, both moving in temperature as
well as moving in magnetic field, between a spin glass behavior and a
paramagnetic one. We have a picture of the phase
diagram composed by three points (using the notation $(T,h)$):  
$(1.138,0)$, $(\sim 1.1,0.2)$ and $(0.714,\sim 0.6)$ 
(with the symbol $\sim$ we denote that the figure
that follows is only indicative). In addition we know that for the
Gaussian model there is a critical point at $(0,0.6)$, but this
critical  magnetic field should be modified to take into account that
we are using binary couplings.

Finally we have checked that the overlap probability distribution of
our model, $P(q)$, (obtained via the static-dynamics link: $X(C) \to
x(q)$), which is not trivial (at least within our numerical
precision, see section II.), satisfies the PaT Ansatz.

\section{Acknowledgments}

We thank partial financial support from CICyT (AEN97-1680,
FPA2000-0956, FPA2001-1813 and PB98-0842) and DGA (P46/97).  We also
want to acknowledge the contribution of Prof. Domingo Gonz\'alez to
the development of spanish Physics, and in particular his help to our
group, which has been essential to build our dedicated computers.  We
finally thank Giorgio Parisi for interesting comments. S. Jimenez is a
DGA fellow.

\end{document}